\documentstyle[preprint,eqsecnum,aps,epsf]{revtex}

\begin{document}
\draft

\title{Polarized Structure Function $g_2$ in the CM bag model\\}

\author{X. Song$^*$\\
{\it Institute of Nuclear and Particle Physics,}\\ 
{\it Department of Physics, University of Virginia,} \\
{\it Charlottesville, VA 22901, USA.}}

\maketitle

\begin{abstract}
The spin-dependent structure functions $g_1(x)$, 
$g_2(x)$, ${g}_2^{WW}(x)$ and ${\bar g}_2(x)$ 
and their moments are studied in the CM bag model. 
The results show that (i) $\int_0^1g_2(x)dx=0$, 
i.e. the Burkhardt-Cottingham sum rule holds, hence 
$g_2(x)$ must have at least one non-trivial zero 
besides $x=0$ and $x=1$. (ii) $\int_0^1x^2g_2(x)dx$ 
is negative for the proton, neutron and deuteron. 
(iii) $\int_0^1x^2g_2(x)dx$ is about one order of
magnitude smaller than $\int_0^1x^2g_1(x)dx$, hence
the twist-3 matrix element is approximately equal 
to the twist-2 matrix element. The results are 
compared with most recent data and predictions 
from the MIT bag model, lattice QCD and QCD sum 
rules. 

\end{abstract}

\bigskip

\pacs{13.60.Hb; 14.39.Ba\\
$^*$ E-mail address: xs3e@virginia.edu\\}

\widetext

\section{Introduction}
 
In recent years, the investigation of the spin 
structure of the nucleon by using deep inelastic 
lepton-hadron scattering (DIS) experiments has been 
an exciting and controversial field in hadron physics. 
The experiments performed by EMC, SMC at CERN 
\cite {EMC,SMC}, and E142, E143 at SLAC 
\cite {E142,E143a,E143b} provide direct 
information on the matrix elements of spin-dependent 
operators in the nucleon. The spin-dependent DIS 
cross section is determined by the antisymmetric 
part of the hadronic tensor
$$W_{\mu\nu}^{A}=i\epsilon_{\mu\nu\sigma\rho}
{{q^{\rho}}\over {\nu}}
\{ S^{\sigma}g_1(x,Q^2)+[S^{\sigma}-
P^{\sigma}{{S\cdot q}\over {P\cdot q}}]g_2(x,Q^2)\}
\eqno (1)$$
where $P$ and $q$ are the four vectors of the nucleon 
and virtual photon momentums. $g_1(x,Q^2)$ and $g_2(x,Q^2)$ 
are two spin-dependent structure functions for the Bjorken 
variable $x=Q^2/2P\cdot q\equiv Q^2/2M\nu$ with $Q^2=-q^2$ 
is the transfered four momentum squared and 
$S^{\sigma}={\bar U}(P,S)\gamma^{\sigma}\gamma_5U(P,S)$ 
is the covariant spin vector of the nucleon.

According to Operator Product Expansion (OPE) analysis, 
to order $M^2/Q^2$, the lowest two moments of $g_1(x,Q^2)$ 
and $g_2(x,Q^2)$ can be written (QCD radiative corrections 
are not included) \cite{sv82a,jaffe90,mit91,ji94,ems94}
$$\int_0^1g_1(x,Q^2)dx={1\over 2}a^{(0)}(Q^2)+{{M^2}\over 
{9Q^2}}[a^{(2)}(Q^2)+4d^{(2)}(Q^2)+4f^{(2)}(Q^2)]
\eqno(2a)$$
$$\int_0^1g_2(x,Q^2)dx=0
\eqno(2b)$$
$$\int_0^1x^2g_1(x,Q^2)dx={1\over 2}a^{(2)}(Q^2)+O({{M^2}
\over {Q^2}})
\eqno(2c)$$
$$\int_0^1x^2g_2(x,Q^2)dx=-{1\over 3}a^{(2)}(Q^2)+
{1\over 3}d^{(2)}(Q^2)+O({{M^2}\over {Q^2}})
\eqno (2d)$$
where $a^{(0,2)}(Q^2)$, $d^{(0,2)}(Q^2)$ and $f^{(2)}(Q^2)$ 
depend on the nucleon forward matrix elements of twist-2,
twist-3 and twist-4 operators respectively. For example 
$$a^{(0)}=\sum\limits_{q_f=u,d,s...}e_f^2\Delta {q_f},
\qquad <P,S|{\bar\psi}_f\gamma_{\mu}\gamma_5\psi_f|P,S>
\equiv 2\Delta {q_f}S_{\mu}
\eqno (3) $$
$\Delta {q_f}$ ($q_f=u,d,s,..$) are axial charges defined 
by the above axial-vectorial matrix elements. The singlet 
axial charge is proportional to the total quark helicity 
$\Delta\Sigma=\Delta u+\Delta d+\Delta s$ in the nucleon. 
Using the semileptonic weak decay data, which are related 
to nonsinglet axial charges, the 1988 EMC result seemed 
to indicate that $\Delta\Sigma$ is surprisingly small and 
led to so-called ``spin crisis''. Since then, an intensive 
study of $g_1$ has been conducted. The efforts both from 
experimental and theoretical works led to a deeper 
understanding of internal spin structure of the nucleon, 
although many questions remain. The most recent reviews on 
this subject can be found in \cite{jaffe95,ans94,close}.
Neglecting the terms of order $M^2/Q^2$ in eqs.(2a-d), 
the longitudinal polarized structure function $g_1(x,Q^2)$ 
receives only twist-2 contributions. On the other hand, 
the structure function $g_2(x,Q^2)$ contains not only 
twist-2 but also twist-3 contributions corresponding to
the matirx element $d^2(Q^2)$. The twist-3 
contributions coming from spin-dependent quark
gluon correlations do not vanish even in the large $Q^2$ 
limit \cite{jaffe90,hey72,ji93,kod}. 

To separate the twist-2 and twist-3 contributions of
$g_2(x,Q^2)$, one can write (Wandzura-Wilczek\cite {ww}) 
$$g_2(x,Q^2)=g_2^{WW}(x,Q^2)+{\bar g}_2(x,Q^2)
\eqno (4)$$
where the twist-2 piece (see, however, discussion in 
section III) is 
$$g_2^{WW}(x,Q^2)=-g_1(x,Q^2)+\int_x^1{{dy}\over y}g_1(y,Q^2)
\eqno (5a)$$ 
and the twist-3 piece is 
$${\bar g}_2(x,Q^2)=g_1(x,Q^2)+g_2(x,Q^2)-\int_x^1
{{dy}\over y}g_1(y,Q^2)
\eqno (5b)$$ 
It is easy to check that, to order O($M^2/Q^2)$, 
we have
$$\int_0^1g_2^{WW}(x,Q^2)dx=0,\qquad 
\int_0^1x^2g_2^{WW}(x,Q^2)dx=-{1\over 3}a^{(2)}
\eqno (6a)$$
$$\int_0^1{\bar g}_2(x,Q^2)dx=0,\qquad 
\int_0^1x^2{\bar g}_2(x,Q^2)dx={1\over 3}d^{(2)}
\eqno (6b)$$
hence the twist-2 and twist-3 contributions of $g_2(x,Q^2)$ 
are separated. Measuring $g_1(x,Q^2)$ and $g_2(x,Q^2)$ as 
functions of $x$ and $Q^2$, one can obtain $g_2^{WW}(x,Q^2)$, 
$\bar g_2(x,Q^2)$ and their third and even higher moments. 
Up to order $M^2/Q^2$, $g_2(x,Q^2)$ uniquely measures the 
twist-3 contributions without involving any model-dependent 
analysis. Hence $g_2$ provides more detailed information 
about nucleon structure than does $g_1$. Very recently, a 
preliminary experimental result of $g_2$ has been published 
\cite{SMC94} and more precise data obtained by E143 experiment 
\cite{E143c} at SLAC should be published soon. Since 
$g_2$ was discussed only briefly in a previous paper 
\cite{song94}, it is appropriate to give a detailed analysis 
in the modified Center-of-Mass (CM) bag model, using newly 
obtained data on $g_2$. In section II, the CM bag model is
briefly reviewed and the results for $g_2(x)$, ${g}_2^{WW}(x)$ 
and ${\bar g}_2(x)$, and their moments $a^{(0,2)}(Q^2)$, 
$d^{(0,2)}(Q^2)$ are presented. Discussion and comparison 
with most recent data and other model results are given in 
section III. A brief summary is given in section IV.

\section{CM bag model}

As mentioned in a previous paper \cite{song94}, the 
bag model does not contain gluon fields explicitly, but the 
boundary of the bag-confined quarks simulates the binding 
effect coming from quark-gluon and gluon-gluon interactions 
(the gluon contribution to the proton spin in the bag model 
has been discussed by Jaffe \cite{jaffe95p} recently). 
Hence, the structure function $g_2$ calculated in the bag 
model does include higher twist effects. We have 
reported the CM bag model results \cite{song94} for the 
unpolarized and polarized structure functions $F_1$, $F_2$, 
and $g_1$, and briefly for $g_2$. In this paper, we will 
focus our attention on $g_2$. For reader's convenience, we 
give a brief review for CM bag model calculation of the 
structure functions. One needs to calculate the
hadron tensor
$$W_{\mu\nu}(P,q,S)={1\over {4\pi}}\int d^4ye^{iqy}
<P,S|[J_{\mu}(y),J_{\nu}(0)]|P,S>
$$
and separate the antisymmetric part to obtain
$W_{\mu\nu}^A$ which can be expressed in general as
$W_{\mu\nu}^{A}=i\epsilon_{\mu\nu\sigma\rho}
({{q^{\rho}}/{\nu}})I^{\sigma}(x, Q^2)$ in the Bjorken
limit, where $I^{\sigma}(x,Q^2)$ depends on the model 
of the nucleon and the approximations used in the 
calculation. 

The basic assumptions and approximations of the CM 
bag model are: (i) the nucleon electromagnetic 
current $J_{\mu}$ (or $\gamma$NN vertex, see (2.7) in 
\cite{song94}), can be approximately expressed by 
incoherent sum of single quark electromagnetic currents. 
It implies that the virtual photon interacts with only 
one quark at a time and the other two quarks are spectators; 
this is an impulse approximation. The current includes
not only the contribution of the struck quark but also 
those of the spectator quarks. Since the current satisfies 
translational invariance, four momentum is conserved. 
(ii) the nucleon consists of three valence quarks in their 
$S$-wave state; higher excited states and higher Fock states 
which include gluons and sea quark pairs in addition to three 
valence quarks are neglected. (iii) 
SU(3)$_{flavor}\otimes$SU(2)$_{spin}$ wave functions 
for the proton and neutron are used. Symmetry-breaking effect 
is described in terms of a parameter $\xi\equiv R_d^P/R_d^p<1$, 
which simulates the smaller spatial size for the scalar $u-d$ 
quark pair than that for the vector $u-u$ and $d-d$ quark 
pairs in the nucleon \cite{song92}. (iv) The effect
of quark confinement due to nonperturbative quark-gluon
and gluon-gluon interactions is described in terms of 
bound-state quark spatial wave functions, for instance 
the quark bag wave function in the cavity approximation 
in the MIT bag model or Gaussian-type quark wave function
used in some other models. All necessary 
formulae for the CM bag model calculation can be found in 
\cite{song94}. A formal and general discussion on the 
theoretical basis of the CM bag model has been given in 
\cite{wang95}. We note that in addition to the CM 
bag model calculation, the transverse spin structure 
functions $g_2$ has also been computed in the original 
MIT bag model by Jaffe and Ji \cite{mit91}, and other 
modified versions of the MIT bag model by Schreiber, Signal
and Thomas (SST bag model \cite{sst91}), and by Stratmann 
(MOD model \cite{strat93}). 

Experimentally, $g_1(x,Q^2)$ and $g_2(x,Q^2)$ are 
measured by combining two different cases of deep 
inelastic scattering of polarized leptons on polarized 
nucleons: (i) the beam and target spin orientations are 
parallel, and (ii) the beam and target spin orientations 
are perpendicular. Experimentally, 
$W_{\mu\nu}^{A}=i\epsilon_{\mu\nu\sigma\rho}(q^{\rho}/{\nu}) 
I^{\sigma}(x,Q^2)$ can be calculated from various models of 
the nucleon. In this case it is convenient to choose 
suitable projection operators to extract $g_1$ and 
$g_2$ from model results of $I^{\sigma}(x,Q^2)$. One of 
possible projections is to extract $g_1\equiv g_L$ 
and $g_1+g_2\equiv g_T$ by choosing the nucleon spin 
parallel (`L') or perpendicular (`T') to the 
virtual photon momentum as we did in \cite{song94}. 
It should be noted that it is not necessary to choose 
the same projection as those used in the experimental
analysis.

Several parameters have been used in the CM bag model 
calculation. They are: (a) `bag' radius R=5 GeV$^{-1}$, 
(b) SU(3) symmetry breaking parameter $\xi=0.85$ and 
(c) maximum momentum of quarks inside the nucleon 
$|{\bf p}_{max}|=0.6$ GeV/c. $R$ and $\xi$ were 
determined from the fit of the $rms$ radius of the 
neutron and proton, and the ratio $\mu_n/\mu_p$ 
\cite{song92}. The model with these two parameters 
gives a fairly good result for the electromagnetic 
form factors of the nucleon and the magnetic moments 
of octet baryons. In particular, the neutron charge 
form factor is well reproduced. For the DIS parton 
distributions, the third parameter $|{\bf p}_{max}|
=0.6$ GeV/c has been introduced. It constrains the 
unpolarized valence quark distributions to satisfy 
the sum rules 
$$\int_0^1u_v(x)dx=2,\qquad \int_0^1d_v(x)dx=1 
\eqno (7)$$
As mentioned in \cite{song94}, to compare the model 
results with data, the QCD evolution technique has 
to be used to evolve the parton distributions from the  
renormalization scale $Q_0^2$ to higher $Q^2$ where 
the experiments are performed. We choose $Q_0^2$=0.81 
(GeV/c)$^2$ and QCD scale parameter $\Lambda$=0.3 (GeV/c). 
The CM bag model with these parameters gives a good 
description for both unpolarized and polarized structure 
functions at $0.3<x<1$, where the valence quark 
contributions dominate. For small-$x$ region, $0<x<0.3$, 
the sea quark contributions are necessary. Using some QCD 
inspired phenomenological sea distributions (see 
(4.19) in \cite{song94}), we found that the first moment 
of $g_1^p(x,Q^2)$ is consistent with the experimental 
value and the first moment of $g_1^p(x,Q^2)-g_1^n(x,Q^2)$ 
satisfies the Bjorken sum rule. For the higher moments 
($n\geq 2$), the sea contributions coming mainly from 
the small $x$ region are highly suppressed by the factor 
$x^n$ and are thus less important. Hence we neglect 
possible sea contributions for the higher moments of $g_2$, 
$g_2^{WW}$ and ${\bar g}_2$.

For QCD evolution of the structure function $g_1$ and
twist-2 piece $g_2^{WW}$ from $Q_0^2$ to $Q^2>Q_0^2$, 
the ordinary Altarelli-Parisi equations \cite{apsong} 
can be used. For twist-3 part of $g_2$, however, it 
has been shown \cite{sv82b} that due to mixing of 
twist-3 quark operators and quark-gluon operators 
with same twist and quantum numbers, the number of 
independent operators contributing to ${\bar g}_2$ 
increases with $n$, where $n$ refers to the $n$-th 
moment. It implies that one cannot write down an 
Altarelli-Parisi type evolution equation for $g_2$. 
This feature has been confirmed by several later 
calculations in \cite{bkl83,rat86,bb88,jichou90,ali91} 
and most recently in \cite{kyu95}. Hence there is no 
simple evolution equation for $g_2$ in the general 
case. One has to look for some approximate solutions. 
Two approximate evolution approaches under the limits 
$N_c\rightarrow \infty$ or power $n\rightarrow \infty$ 
were suggested by Ali, Braun and Hiller \cite{ali91}. 
We use the approach in the large $N_c$ limit rather 
than the approximation in the large $n$ limit, which 
only provides the asymptotic behavior of $g_2(x,Q^2)$ 
in the region $x\rightarrow 1$. We also note that as 
far as the third moments $\int_0^1x^2g_2(x,Q^2)dx$
are concerned, the $Q^2$ evolution is straightforward,
i.e. a single power behavior of $ln Q^2$ (for instance
see \cite{kyu95}).

In Fig.1$-$5, we present the results of $g_1^p(x,Q^2)$, 
$g_2^p(x,Q^2)$, $x^2g_2^p(x,Q^2)$,  $g_2^{pWW}(x,Q^2)$ 
(which is determined by $g_1^p(x,Q^2)$), and 
$x^2{\bar g}_2^p(x,Q^2)$ respectively. For the deuteron 
target, $g_1^d(x,Q^2)$, $x^2g_2^d(x,Q^2)$ and 
$x^2{\bar g}_2^d(x,Q^2)$ are shown in Fig. 6-8. All 
theoretical curves are calculated in the CM bag model 
at $Q_0^2$=0.81 (GeV/c)$^2$ and evolved to $Q^2$=5.0 
(GeV/c)$^2$ except for Fig. 4 and Fig. 6, where $Q^2$=4.0 
(GeV/c)$^2$ and $Q^2$=3.0 (GeV/c)$^2$ respectively. 
Comparisons of our results for the third moments of $g_1$ 
and $g_2$ with recent data and other model predictions 
are listed in Tables I, II and III. The data for $g_1$
are taken from \cite{EMC,SMC,E143a,tj95} and those for
$g_2$ are taken from \cite{E143c}.

\section{Results and discussion}

{\bf 1}. For the leading term of first moments of 
$g_1(x)$, we obtain
$$ a_{proton}^{(0)}=0.252,\qquad 
a_{neutron}^{(0)}=-0.112,\qquad
a_{deuteron}^{(0)}=0.064
\eqno (8)$$
which can be compared with SMC data \cite{SMC} at 
$<Q^2>$=5 (GeV/c)$^2$:
$ a_{proton}^{(0)}=0.252\pm 0.036$, 
$a_{neutron}^{(0)}=-0.056\pm 0.024$, 
$a_{deuteron}^{(0)}=0.046\pm 0.050$ and 
E143 data \cite{E143a,E143b} at $<Q^2>$=3 
(GeV/c)$^2$: 
$a_{proton}^{(0)}=0.254\pm 0.022$, 
$a_{neutron}^{(0)}=-0.074\pm 0.027$, 
$a_{deuteron}^{(0)}=0.084\pm 0.010$.
 
{\bf 2}. For the leading term of third moments of 
$g_1(x)$, we get
$$ a_{proton}^{(2)}=2.10\cdot 10^{-2},\qquad 
a_{neutron}^{(2)}=-1.86\cdot 10^{-3},\qquad
a_{deuteron}^{(2)}=8.74\cdot 10^{-3}
\eqno (9)$$
while preliminary data at $<Q^2>$=5 
(GeV/c)$^2$ \cite{E143c} show:
$ a_{proton}^{(2)}=(2.42\pm 0.20)\cdot 10^{-2}$
and $a_{deuteron}^{(2)}=(8.0\pm 1.6)\cdot 10^{-3}$.
Comparisons with other models are listed in Tables I 
and III. We note that no $g_2^n$ data are available 
yet. 

{\bf 3}. Since $g_1^p(x,Q^2)$ is always positive in 
the range $0<x<1$, hence $\int_0^1x^2g_1^p(x)dx$ must 
be positive. From eq.(2c), the leading term of 
$\int_0^1x^2g_1^p(x)dx$, i.e. $a_{proton}^{(2)}$ must 
also be positive as shown in eq.(9). However, the 
$g_1^n(x)$ as function of $x$ is mostly negative except 
for large $x$ region, where $g_1^n(x)$ is positive but 
very small. Hence its third moment, or $a_{neutron}^{(2)}$, 
is negative. For the deuteron, since the positive 
contribution from the proton is larger than 
the negative contribution from the neutron in the 
$\int_0^1x^2g_1^d(x)dx$, hence our model predicts a 
positive $a_{deuteron}^{(2)}$ as shown in eq.(9) and 
Table III.

{\bf 4}. As mentioned in section I, in the OPE approach
including twist-2 and twist-3 operators in the presence of
QCD corrections, the Burkhardt-Cottingham \cite{BC} sum 
rule $\int_0^1g_2(x)dx=0$ is known to hold. The CM bag 
model predicts 
$$\int_0^1g_2^p(x)dx=-0.0016,\qquad 
\int_0^1g_2^n(x)dx=-0.0047
\eqno (10a)$$ 
comparing to the numerical values for $\int_0^1g_1^{(p,n)}dx$
in eq.(8), they are numerically consistent with zero. Hence 
in our model, the Burkhardt-Cottingham sum rule is satisfied 
within numerical errors. Most recent data \cite{E143c} give
$$\int_{0.03}^1g_2^p(x)=-0.013\pm 0.028,\qquad
\int_{0.03}^1g_2^n(x)dx=-0.033\pm 0.082
\eqno (10b)$$
which are consistent with zero. It should be noted that 
in the MIT bag model with SU(6) symmetry, $g_2^n(x)$ is 
identically zero by itself. However, the CM bag model 
with SU(6) symmetry breaking effects ($\xi\simeq 
0.85<1$) predicts a nonzero $g_2^n(x)$. 

{\bf 5}. Since $g_2(x)$ is not identically zero in the 
model, the Burkhardt-Cottingham sum rule implies that 
$g_2(x)$ must change its sign at some $x=x_0$, where 
$0<x_0<1$, {\it i.e.} $g_2(x)$ must have at least one
non-trivial zero. The question is whether the $x$-behavior 
of $g_2$ satisfies 
$$ {\rm case}\  1:\qquad g_2(x)<0,\quad {\rm for}\ x<x_0;
\qquad g_2(x)>0,\quad {\rm for}\ x>x_0$$
or just opposite
$$ {\rm case}\  2:\qquad g_2(x)>0,\quad {\rm for}\ x<x_0;
\qquad g_2(x)<0,\quad {\rm for}\ x>x_0$$
For case 1, one has $\int_0^1x^2g_2(x)dx>0$, while for 
case 2, $\int_0^1x^2g_2(x)dx<0$. The CM bag model 
predictions for $\int_0^1x^2g_2^{(p,n,d)}(x,Q^2)dx$ are all 
negative (see Table I, II and III or Fig.3 and Fig.7). 
This implies the $x$-behavior of $g_2(x)$ fits case 2. 
The preliminary data \cite{E143c} seem to favor our 
predictions (see, for instance, Fig.2 for the proton). 

{\bf 6}. For both the proton and deuteron, we now 
have $a^{(2)}>0$ and $d^{(2)}-a^{(2)}<0$. In the CM 
bag model, the moment $\int_0^1x^2g_2(x)dx$ ($\simeq
[d^{(2)}-a^{(2)}]/3$) is about one order of magnitude 
smaller than $\int_0^1x^2g_1(x)dx$ ($\simeq a^{(2)}/2$).  
It implies $|d^{(2)}-a^{(2)}|<<a^{(2)}$ (recall $|g_2|<<g_1$), 
or $d^{(2)}\simeq a^{(2)}$. Hence the twist-3 matrix element 
$d^{(2)}$ approximately equals to the twist-2 matrix element 
$a^{(2)}$ and the sign of $d^{(2)}$ should also be positive 
for the proton and deuteron targets. This agrees with data
\cite{E143c} but disagree with the negative sign predicted 
by the QCD sum rules \cite{qcdsr1,qcdsr2} and quenched lattice 
QCD\cite{lat95}. 

For the neutron, since $\int_0^1x^2g_1^n(x)dx$ is negative 
and much larger than $\int_0^1x^2g_2^n(x)dx$ in magnitude, 
hence $d_{neutron}^{(2)}$ is negative. This negative sign 
is consistent with the results given by other approaches
\cite{qcdsr1,qcdsr2,lat95}. In magnitude, our result 
agrees with that given by the quenched lattice QCD, but 
much less than those given by the QCD sum rules.

{\bf 7}. Our results show $d^{(2)}\simeq a^{(2)}$, i.e. the 
twist-3 contribution is almost the same magnitude as 
twist-2 contribution. From (6a) and (6b), it implies
that the $g_2^{WW}(x)$ and ${\bar g}_2(x)$ have opposite 
sign and approximately same magnitude. They almost 
cancel each other and lead to a very small $g_2(x)$ 
(see also Fig. 12a,b in \cite{song94}). For the same 
reason the original Wandzura-Wilczek relation 
$$g_2(x,Q^2)=-g_1(x,Q^2)+\int_x^1{{dy}\over y}g_1(y,Q^2)
$$ 
is not a good approximation and the higher twist 
contributions may not be neglected. As
pointed by Cortes, Pire and Ralston \cite{cpr92} 
that the original Wandzura-Wilczek relation was
derived by using the Dirac equation for free and
massless quarks. Including quark mass effect and 
gluon dependent term, an extended Wandzura-Wilczek 
relation was given (similar formula without quark
mass term has been given by Jaffe \cite{jaffe90}):
$$g_2(x,Q^2)=-g_1(x,Q^2)+
\int_x^1{{dy}\over y}g_1(y,Q^2)
-{{m_q}\over M}\int_x^1{{dy}\over y}{{\partial h_T}\over {\partial y}} 
-\int_x^1{{dy}\over y}{{\partial \xi}\over {\partial y}} 
$$ 
where the mass dependent term is another twist-2 piece
which is related to `transversity' $h_T$. The last term
is the `true' twist-3 piece arising from quark-gluon
correlation. As emphasized in \cite{cpr92} that when
the quark mass term is included the twist-3 term cannot 
be isolated in a model independent way with a measurement 
of $g_1$ and $g_2$. However, neglecting the strange quark
contribution, the quark mass term ($\sim m_q/M$) should 
be negligible for up and down quarks. Hence the separation
of twist-2 and twist-3 pieces in eq.(4) seems to be a 
reasonable approximation.

{\bf 8}. According to the OPE analysis, neglecting the 
$M^2/Q^2$ corrections, the general formulae for the 
moments of the structure functions are
$$\int_0^1x^ng_1(x,Q^2)={1\over 2}a^{(n)}(Q^2),\qquad 
(n=0,2,4,...)
\eqno (11a)$$
$$\int_0^1x^ng_2(x,Q^2)=-{n\over {2(n+1)}}[a^{(n)}(Q^2)
-d^{(n)}(Q^2)]
\qquad (n=2,4,6,...)
\eqno (11b)$$
for $n=2$, they reduced to (2c) and (2d). From  (11a) and 
(11b), one obtains
$$\int_0^1x^n[g_1(x,Q^2)+{{n+1}\over n}g_2(x,Q^2)]dx
={1\over 2}d^{(n)}(Q^2)\qquad (n=2,4,6,...)
\eqno (12)$$
hence one has for $n=2$
$$d^{(2)}=2\int_0^1x^2[g_1(x,Q^2)+{3\over 2}g_2(x,Q^2)]dx
\eqno (13)$$
The CM bag model prediction for the function 
$x^2[g_1(x,Q^2)+{3\over 2}g_2(x,Q^2)]$ for the proton 
is shown in Fig. 9 and that for the deuteron is shown
in Fig. 10. One can see that in both cases, the model 
predictions for $d^{(2)}$ are nonzero and positive.
This seems to be consistent with recent data \cite{E143c}. 

{\bf 9}. If one assumes that (12) holds also for $n=1$ 
one obtains
$$\int_0^1x[g_1(x,Q^2)+2g_2(x,Q^2)]dx={1\over 2}d^{(1)}(Q^2)
\eqno (14a)$$
To lowest order in $\alpha_s$, it was shown \cite{ans94} by
using the Field Theoretical Parton Model that $d^{(1)}(Q^2)$
vanishes in the chiral limit, one has the 
Efremov-Leader-Teryaev (ELT) sum rule
$$\int_0^1x[g_1(x,Q^2)+2g_2(x,Q^2)]dx=0
\eqno (14b)$$
However, our results of $g_1$ and $g_2$ do not satisfy 
this sum rule. To demonstrate this, we plot 
$g_1^{(p,d)}(x,Q^2)+2g_2^{(p,d)}(x,Q^2)$ as functions 
of $x$ in Fig.11 and Fig. 12 and compare them with 
recent proton and deuteron data \cite{E143c} respectively. 
One can see that our model predictions are consistent with 
the SLAC E143 data. However, the ELT sum rule seems not to 
be supported by the data (at least for the proton data). 

\section{Summary}

The study of transverse spin structure function $g_2(x,Q^2)$  
has both theoretical and experimental interest. In the most
naive parton model, the quark is asymptotic free and has no 
transverse momentum, $g_2(x)$ is identically zero. However, 
quarks inside the nucleon are not free, the binding effect, 
which arising from quark-gluon interactions, causes a nonzero 
transverse momentum for quarks and leads to $g_2\neq 0$. 
Measurements of $g_2$ or $g_T=g_1+g_2$ allow us to get more 
information about binding effects, which are mainly formulated 
as `higher twist effects'. On the other hand, since $Q^2$ is 
large in the deep inelastic scattering, quark binding effects 
should not be significant and $g_2$ should be small, especially
compared to $g_1$. This seems to agree with most recent exprimental 
result \cite{E143c}. Our model calculation is consistent with
this conclusion. It should be noted, however, that the theoretical 
predictions given by different models or approaches seem not to 
fully agree with each other because of different approximations.
In addition, as mentioned in section III that another twist-2
piece which is related to the quark mass and `transversity'
$h_T$ has been neglected in our discussion. On the experimental 
side, the errors of data are still quite large. We hope that 
several new experiments \cite{E155,SMC94a,hera} to be performed 
in the next few years will provide more precise data and tell us 
more about the quark spin (including longitudinal and transverse) 
distributions in the nucleon.

\acknowledgements

I wish to thank X. Ji for useful comments and suggestions, and 
thank P. K. Kabir, J. S. McCarthy and H. J. Weber for helpful 
discussions. I also thank O. Rondon-Aramayo for providing the 
updated E143 data and suggestions. This work has been supported 
by the U.S. Department of Energy and the Institute of Nuclear 
and Particle Physics, University of Virginia, USA. 

\vfill\eject


\vfill\eject

Table I. Comparison of model results with proton data
$$
\offinterlineskip \tabskip=0pt 
\vbox{ 
\halign to 1.0\hsize 
   {\strut
   \vrule#                         
   \tabskip=0pt plus 30pt
 & \hfil #  \hfil                  
 & \vrule#                         
 & \hfil #  \hfil                  
 & \hfil #  \hfil                  
 & \vrule#                         
 & \hfil #  \hfil                  
 & \hfil #  \hfil                  
 & \vrule#                         
 & \hfil #  \hfil                  
 & \hfil #  \hfil\quad             
   \tabskip=0pt                    %
 & \vrule#                         
   \cr                             
\noalign{\hrule}
&&&&&&&&&&&\cr
&proton   &&  $\int_0^1x^2g_1^p(x,Q^2)dx$  & &&
$\int_0^1x^2g_2^p(x,Q^2)dx$  & &&         $d^{(2)}_p$ &  &\cr
&   && =${1\over 2}a^{(2)}_p$  & &&
$={1\over 3}(d^{(2)}_p-a^{(2)}_p)$&&&
&&\cr
&&&&&&&&&&&\cr
\noalign{\hrule}
&&&&&&&&&&&\cr
& This paper && $1.05\cdot 10^{-2}$ &&&  
$-0.12\cdot 10^{-2}$&&&
$1.74\cdot 10^{-2}$ & &\cr
& MIT bag model\cite{mit91,bag95}    && $-$  &&&  
$-$&&& 1.0$\cdot 10^{-2}$ & &\cr
& QCD sum rule\cite{qcdsr1}    && $-$ &&&  
$-$ &&&$-(0.6\pm 0.3)\cdot 10^{-2}$ & &\cr
& QCD sum rule\cite{qcdsr2}    && $-$ &&&  
$-$&&& $-(0.3\pm 0.3)\cdot 10^{-2}$& &\cr
& Lattice QCD\cite{lat95}    &&$(1.50\pm 0.32)\cdot 10^{-2}$ &&&  
$-(2.61\pm 0.38)\cdot 10^{-2}$&&&$-(4.8\pm 0.5)\cdot 10^{-2}$& &\cr
& &&  &&&  &&&  &&\cr
& data\cite{E143c}   && (1.21$\pm$0.10)$\cdot 10^{-2}$ &&& 
$-(0.63\pm 0.18)\cdot 10^{-2}$
 &&& $(0.54\pm 0.50)\cdot 10^{-2}$  &&\cr
&&&&&&&&&&&\cr
\noalign{\hrule}
}}$$

\smallskip

Table II. Comparison of model results for the neutron 
$$
\offinterlineskip \tabskip=0pt 
\vbox{ 
\halign to 1.0\hsize 
   {\strut
   \vrule#                         
   \tabskip=0pt plus 30pt
 & \hfil #  \hfil                  
 & \vrule#                         
 & \hfil #  \hfil                  
 & \hfil #  \hfil                  
 & \vrule#                         
 & \hfil #  \hfil                  
 & \hfil #  \hfil                  
 & \vrule#                         
 & \hfil #  \hfil                  
 & \hfil #  \hfil\quad             
   \tabskip=0pt                    %
 & \vrule#                         
   \cr                             
\noalign{\hrule}
&&&&&&&&&&&\cr
&neutron   && $\int_0^1x^2g_1^n(x,Q^2)dx$  & &&
$\int_0^1x^2g_2^n(x,Q^2)dx$  & &&         $d^{(2)}_n$&  &\cr
&   && =${1\over 2}a^{(2)}_n$  & &&
$={1\over 3}(d^{(2)}_n-a^{(2)}_n)$ &&&
 &&\cr
&&&&&&&&&&&\cr
\noalign{\hrule}
&&&&&&&&&&&\cr
& This paper  && $-0.93\cdot 10^{-3}$ &&& 
$-0.23\cdot 10^{-3}$ &&&$-2.53\cdot 10^{-3}$&&\cr
& MIT bag model\cite{mit91,bag95} && $-$ &&& 
$-$ &&& 0  &&\cr
& QCD sum rule\cite{qcdsr1}   && $-$ &&& 
$-$&&&$-(30\pm 10)\cdot 10^{-3}$&&\cr
& QCD sum rule\cite{qcdsr2}   && $-$ &&& 
$-$&&&$-(25\pm 10)\cdot 10^{-3}$&&\cr
& Lattice QCD\cite{lat95}   && $-(1.2\pm 2.0)\cdot 10^{-3}$ &&& 
$-(0.4\pm 2.2)\cdot 10^{-3}$ &&&
$-(3.9\pm 2.7)\cdot 10^{-3}$&&\cr
& &&  &&&  &&&  &&\cr
& data   &&$-$  &&&$-$ 
 &&&$-$  &&\cr
&&&&&&&&&&&\cr
\noalign{\hrule}
}}$$

Table III. Comparison of model results with deuteron data
$$
\offinterlineskip \tabskip=0pt 
\vbox{ 
\halign to 1.0\hsize 
   {\strut
   \vrule#                         
   \tabskip=0pt plus 30pt
 & \hfil #  \hfil                  
 & \vrule#                         
 & \hfil #  \hfil                  
 & \hfil #  \hfil                  
 & \vrule#                         
 & \hfil #  \hfil                  
 & \hfil #  \hfil                  
 & \vrule#                         
 & \hfil #  \hfil                  
 & \hfil #  \hfil\quad             
   \tabskip=0pt                    %
 & \vrule#                         
   \cr                             
\noalign{\hrule}
&&&&&&&&&&&\cr
&deuteron   && $\int_0^1x^2g_1^d(x,Q^2)dx$  & &&
$\int_0^1x^2g_2^d(x,Q^2)dx$  & &&         $d^{(2)}_d$&  &\cr
&   && =${1\over 2}a^{(2)}_d$  & &&
$={1\over 3}(d^{(2)}_d-a^{(2)}_d)$ &&&
&&\cr
&&&&&&&&&&&\cr
\noalign{\hrule}
&&&&&&&&&&&\cr
& This paper  && 4.37$\cdot 10^{-3}$ &&& 
$-0.65\cdot 10^{-3}$ &&& $6.79\cdot 10^{-3}$  &&\cr
& MIT bag model\cite{mit91,bag95} && $-$ &&& 
$-$ &&& $5.0\cdot 10^{-3}$  &&\cr
& QCD sum rule\cite{qcdsr1}   && $-$ &&& 
$-$ &&&$-(17\pm 5)\cdot 10^{-3}$ &&\cr
& QCD sum rule\cite{qcdsr2}   && $-$ &&& 
$-$ &&&$-(13\pm 5)\cdot 10^{-3}$  &&\cr
& Lattice QCD\cite{lat95}   && $(6.9\pm 2.6)\cdot 10^{-3}$ &&& 
$-(13.3\pm 3.0)\cdot 10^{-3}$ &&&$-(22\pm 6)\cdot 10^{-3}$  &&\cr
& &&  &&&  &&&  &&\cr
& data\cite{E143c}   && (4.0$\pm 0.8)\cdot 10^{-3}$ &&& 
$-(1.4\pm 3.0)\cdot 10^{-3}$ &&& $(3.9\pm 9.2)\cdot 10^{-3}$ &&\cr
&&&&&&&&&&&\cr
\noalign{\hrule}
}}$$

\vfill\eject

\vspace{0.2 cm}


\begin{figure}[h]
\epsfxsize=5.0in
\centerline{\epsfbox{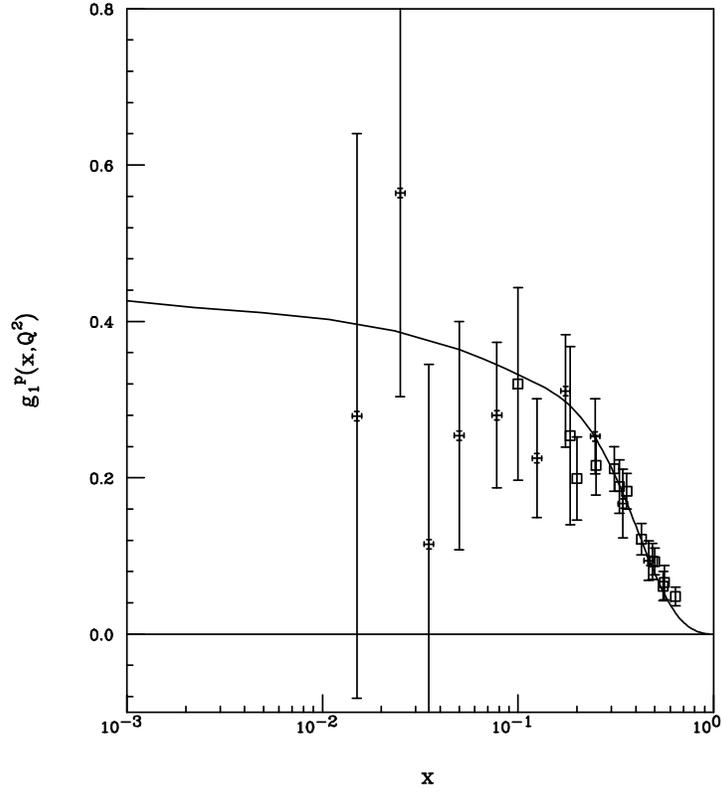}}
\caption{CM bag model result [20] for $g_1^p(x,Q^2)$ 
at $Q^2$=5.0 (GeV/c)$^2$, data from [1,2,4,34].}
\end{figure}

\begin{figure}[h]
\epsfxsize=5.0in
\centerline{\epsfbox{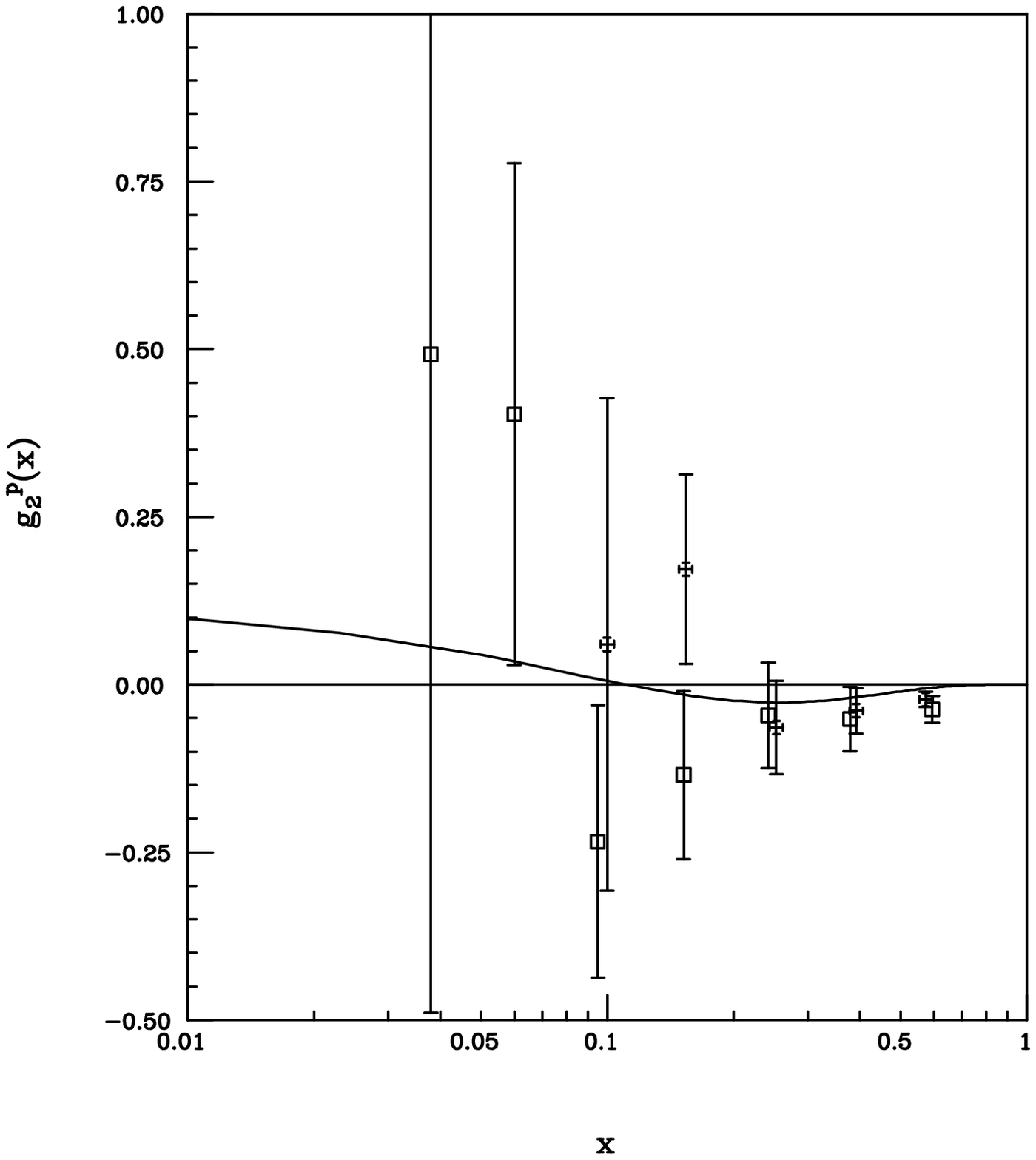}}
\caption{CM bag model prediction for $g_2^p(x,Q^2)$ 
at $Q^2$=5.0 (GeV/c)$^2$, data from [19].}
\end{figure}

\begin{figure}[h]
\epsfxsize=5.0in
\centerline{\epsfbox{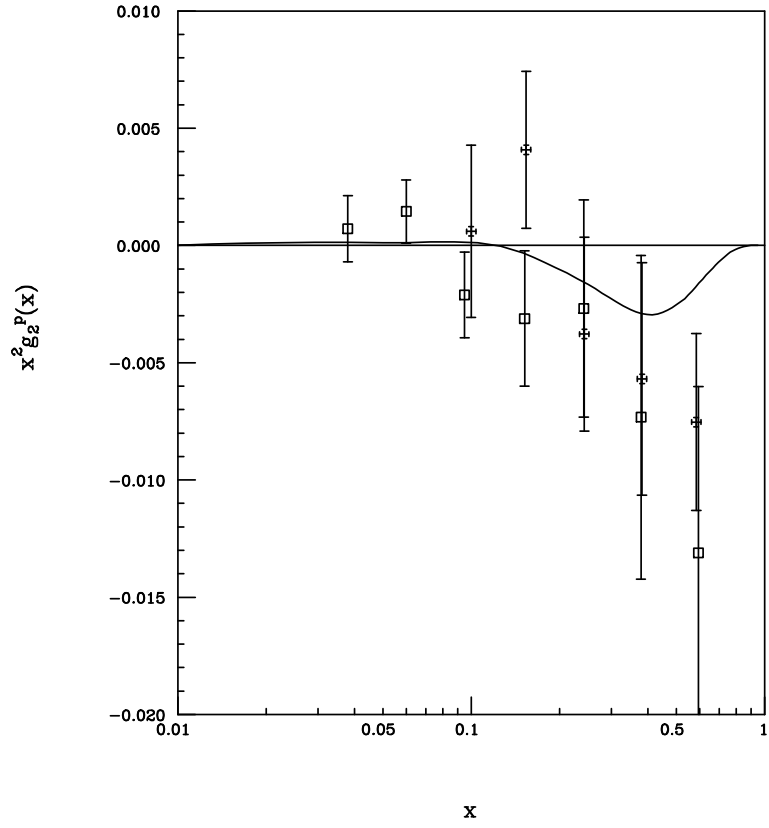}}
\caption{Same as Fig.2, but for $x^2g_2^p(x,Q^2)$ at 
$Q^2$=5.0 (GeV/c)$^2$, data from [19].}
\end{figure}

\begin{figure}[h]
\epsfxsize=5.0in
\centerline{\epsfbox{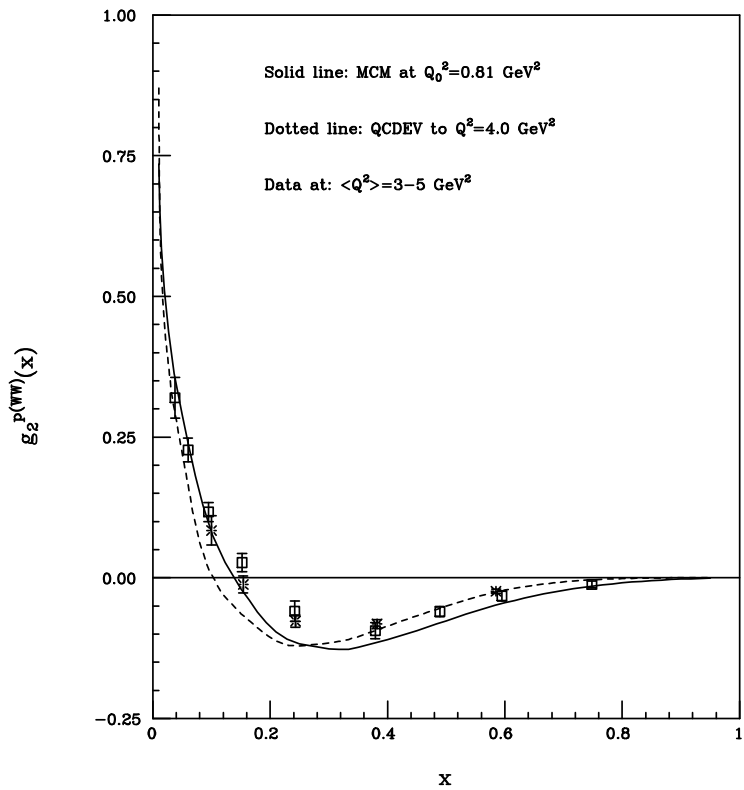}}
\caption{Evolution of twist-2 piece
$g_2^{pWW}(x,Q^2)$ in the CM bag model at $Q^2$=0.81 
and 4.0 (GeV/c)$^2$, data from [19].}
\end{figure}

\begin{figure}[h]
\epsfxsize=5.0in
\centerline{\epsfbox{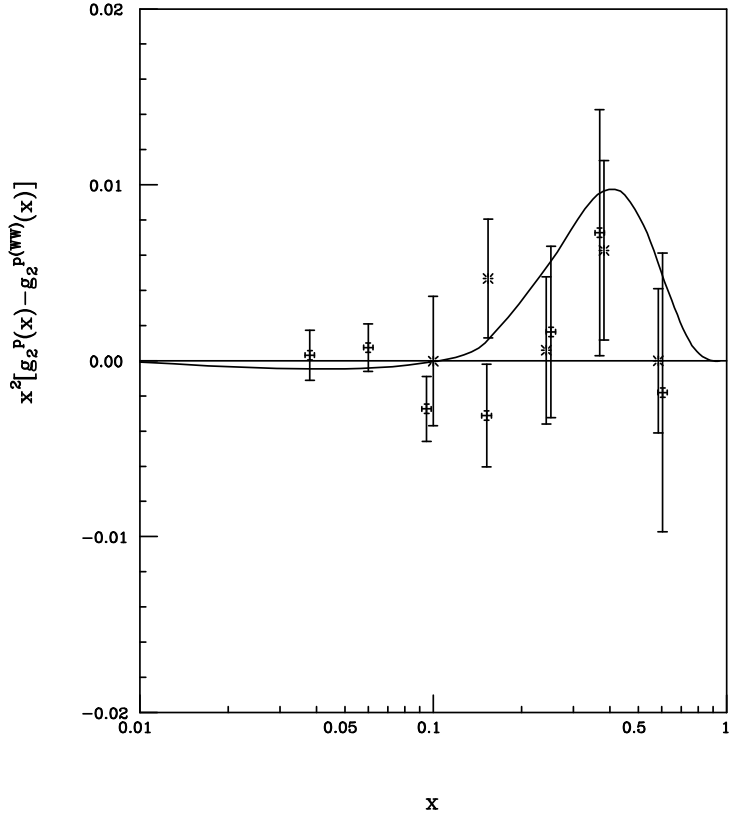}}
\caption{The twist-3 contribution $x^2{\bar g}_2(x,Q^2)$ 
calculated in the CM bag model and evolved to $Q^2$=5.0 
(GeV/c)$^2$, data from [19].}
\end{figure}

\begin{figure}[h]
\epsfxsize=5.0in
\centerline{\epsfbox{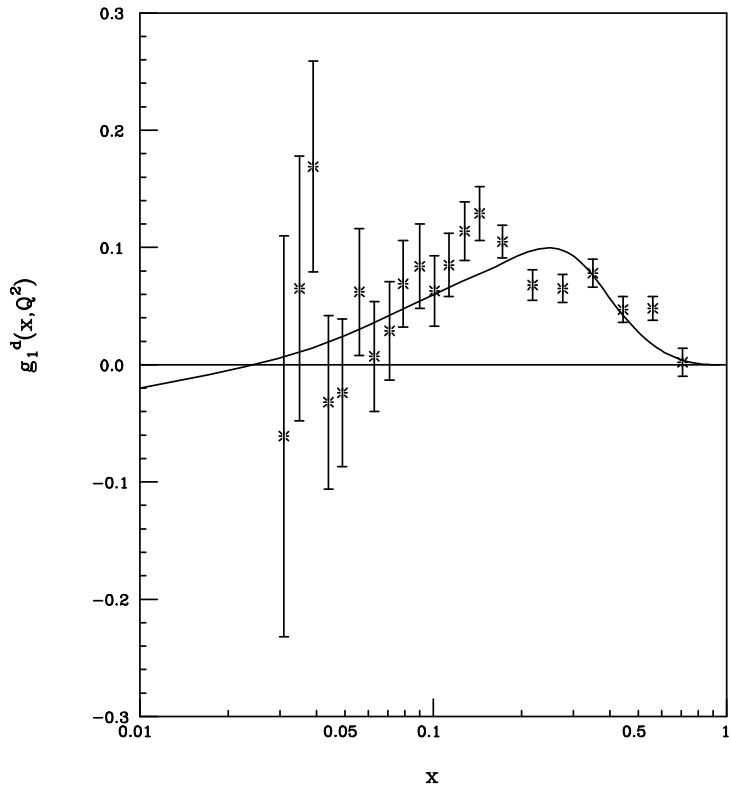}}
\caption{The deuteron structure function $g_1^d(x,Q^2)$ 
calculated in the CM bag model and evolved to $Q^2$=3.0 
(GeV/c)$^2$, data from [34].}
\end{figure}

\begin{figure}[h]
\epsfxsize=5.0in
\centerline{\epsfbox{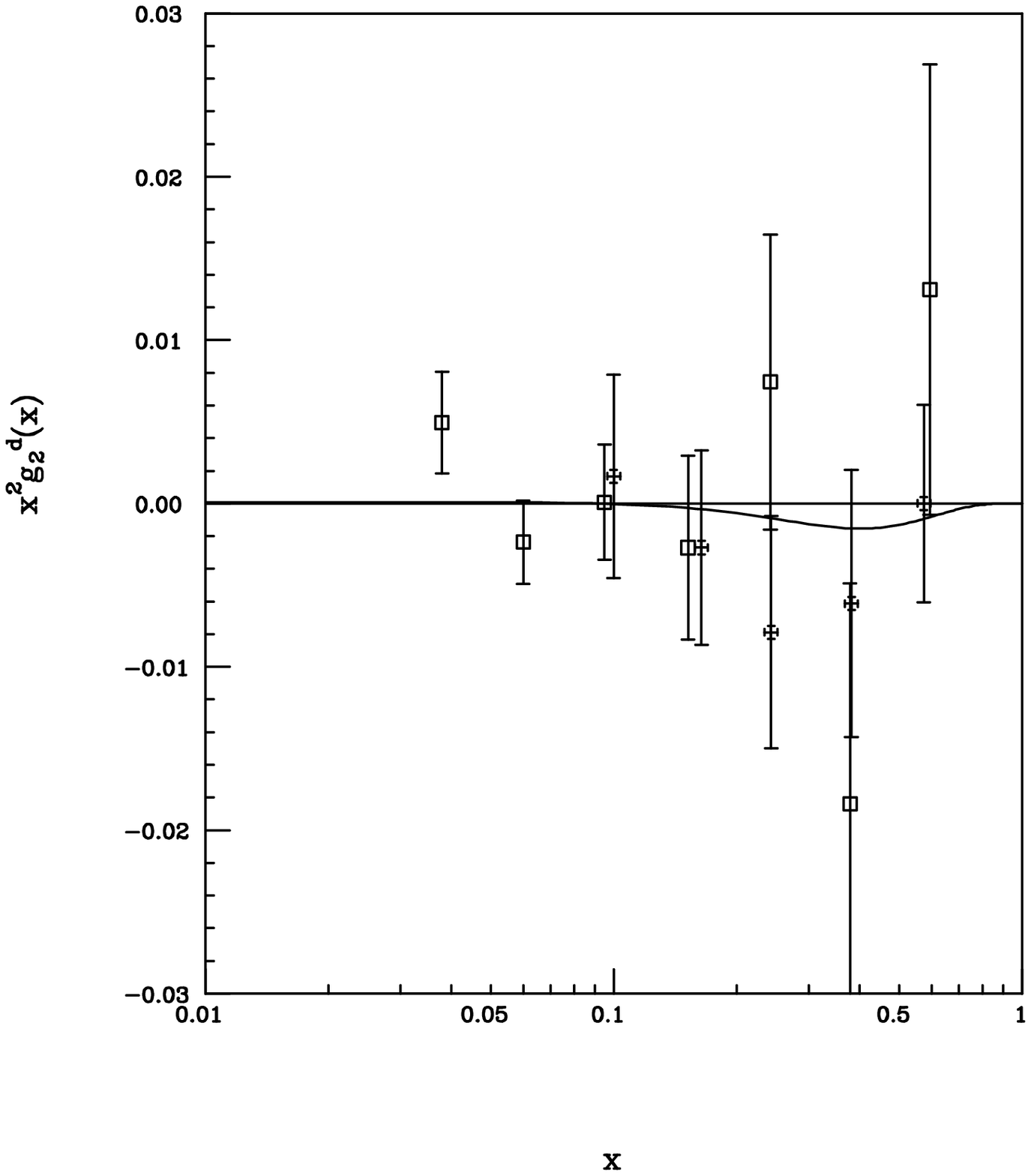}}
\caption{The deuteron transverse
structure function $x^2g_2^d(x,Q^2)$ calculated in the
CM bag model and evolved to 
$Q^2$=5.0 (GeV/c)$^2$, data from [19].}
\end{figure}

\begin{figure}[h]
\epsfxsize=5.0in
\centerline{\epsfbox{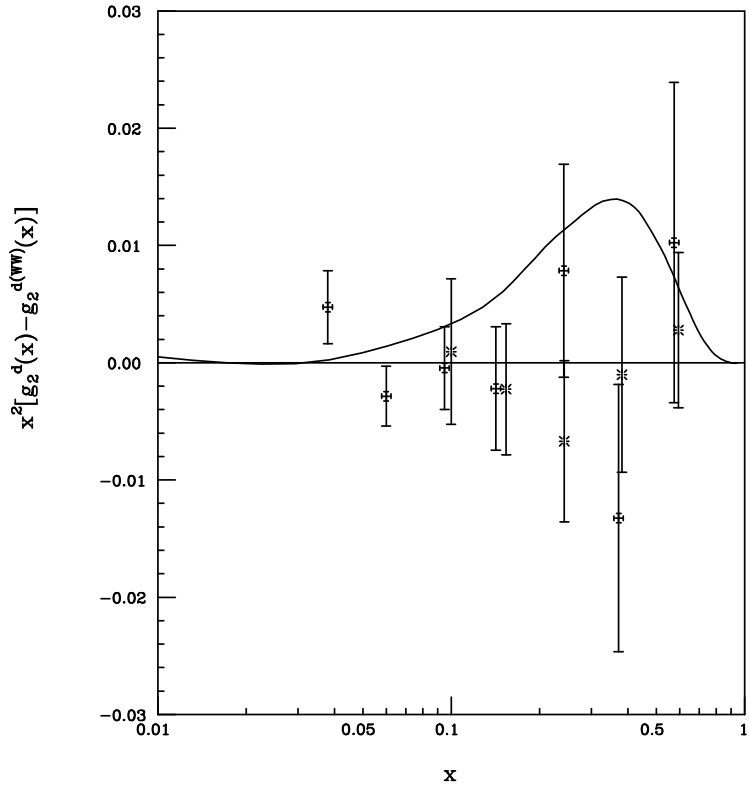}}
\caption{Same as Fig.7 but for deuteron twist-3 piece
$x^2{\bar g}_2^d(x,Q^2)$, data from [19].}
\end{figure}

\begin{figure}[h]
\epsfxsize=5.0in
\centerline{\epsfbox{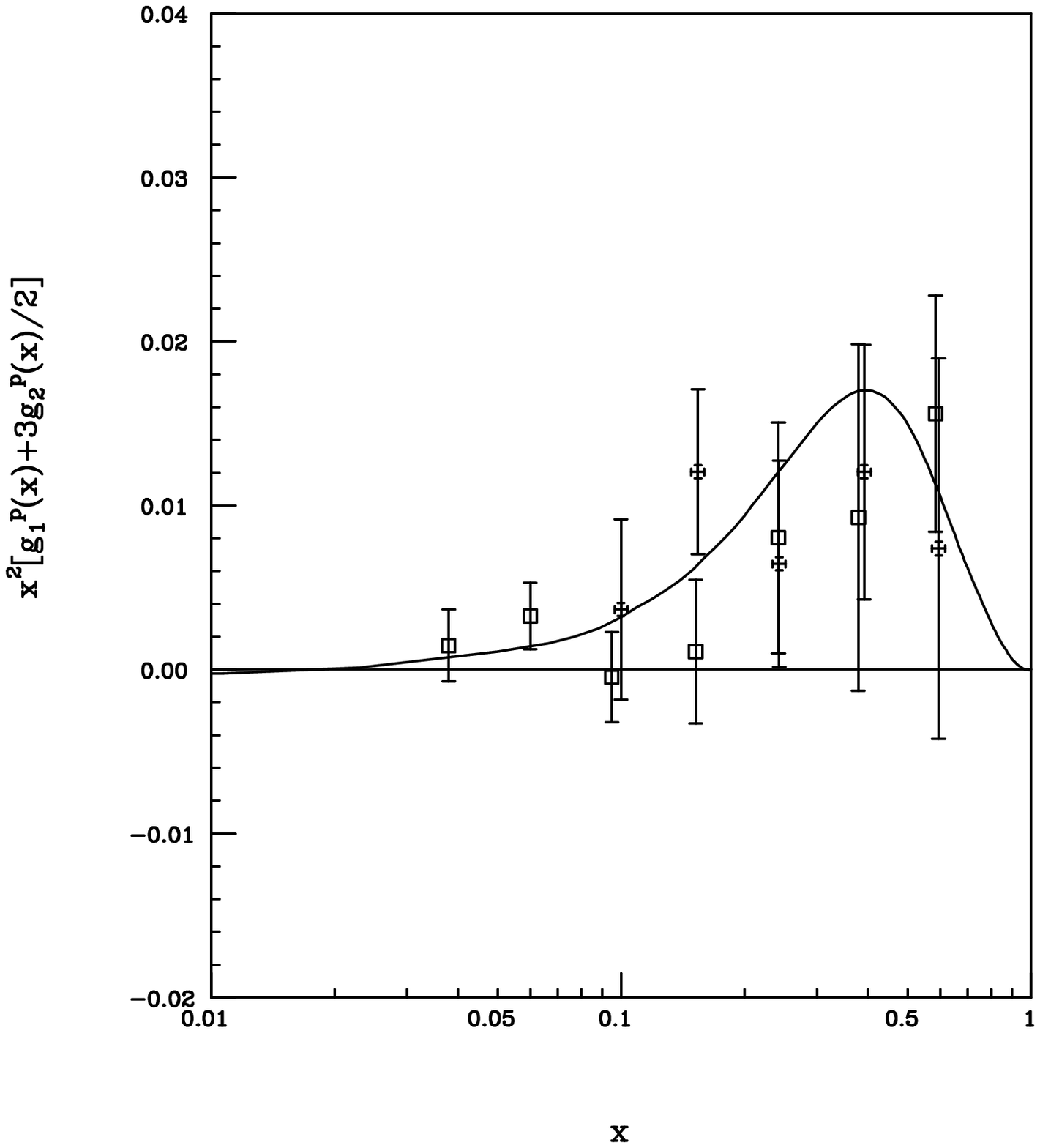}}
\caption{CM bag model prediction for 
$x^2[g_1^p(x,Q^2)+{3\over 2}g_2^p(x,Q^2)]$ at 
$Q^2$=5.0 (GeV/c)$^2$, data from [19].} 
\end{figure}

\begin{figure}[h]
\epsfxsize=5.0in
\centerline{\epsfbox{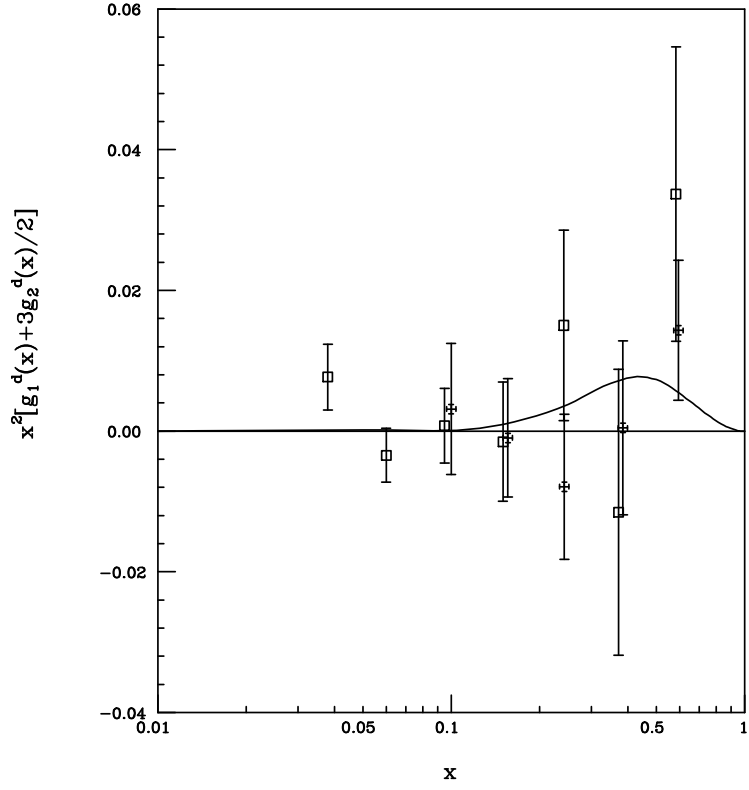}}
\caption{Same as Fig.9 but for $x^2[g_1^d(x,Q^2)+
{3\over 2}g_2^d(x,Q^2)]$ at $Q^2$=5.0 (GeV/c)$^2$. }
\end{figure}

\begin{figure}[h]
\epsfxsize=5.0in
\centerline{\epsfbox{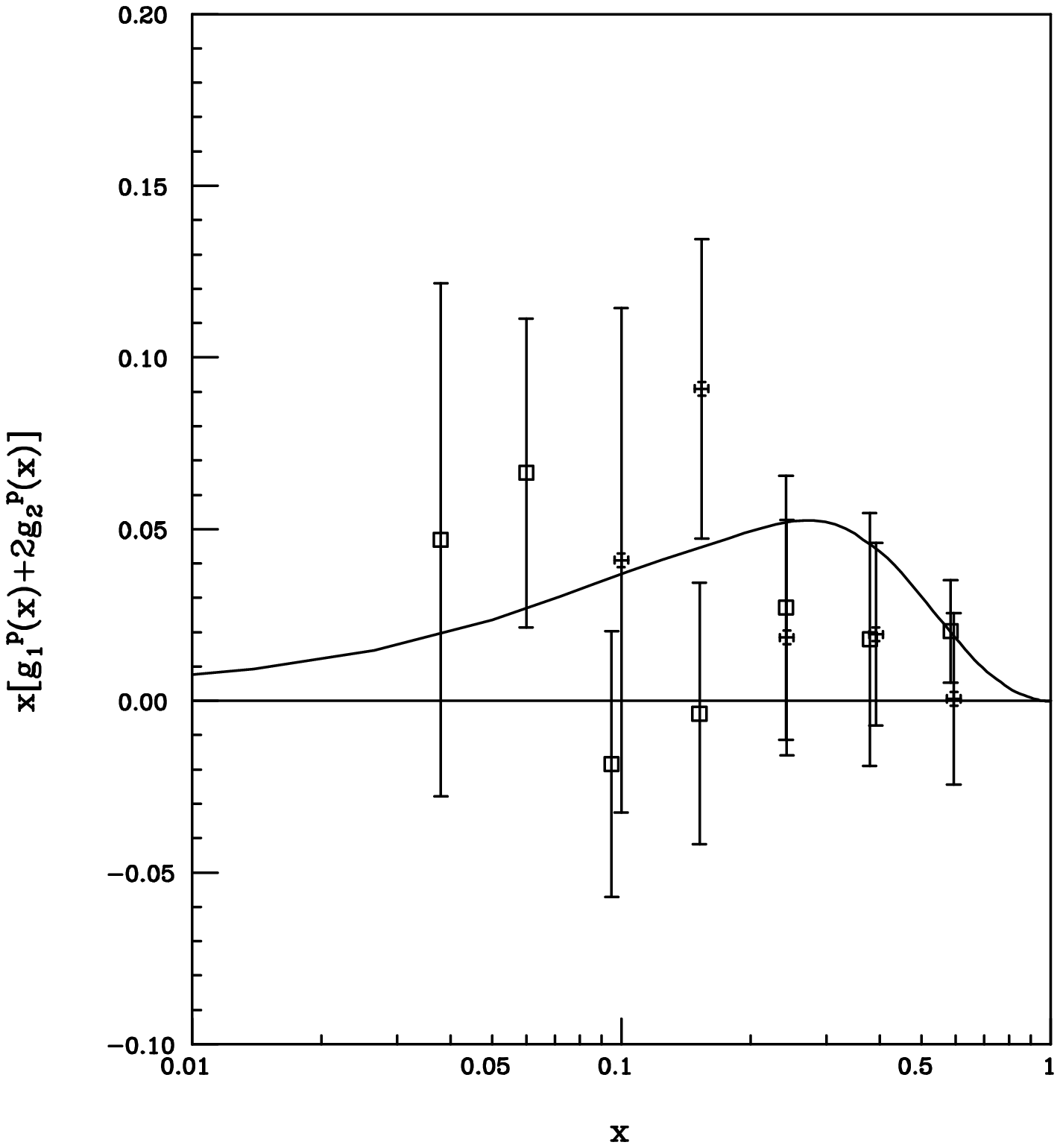}}
\caption{CM bag model prediction for $x[g_1^p(x,Q^2)+
2g_2^p(x,Q^2)]$ at $Q^2$=5.0 (GeV/c)$^2$, data from [19].}
\end{figure}

\begin{figure}[h]
\epsfxsize=5.0in
\centerline{\epsfbox{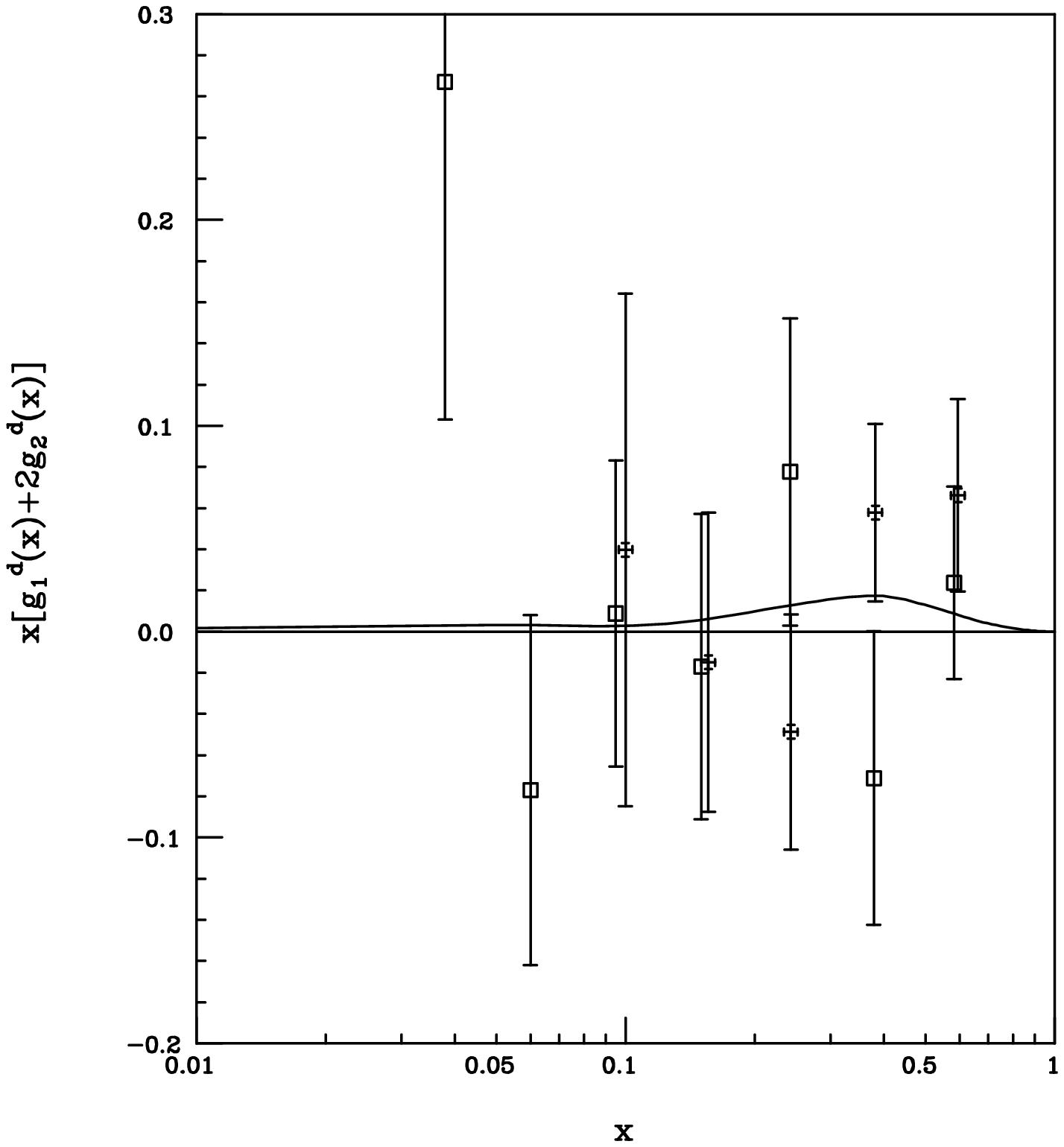}}
\caption{Same as Fig.11 but for the deuteron target, data
from [19].}
\end{figure}


\baselineskip 16pt
\begin{thebibliography}{99}


\bibitem{EMC}
     J.~Ashman {\it et al.} {\sl Phys.Lett.} {\bf B206}, 346 (1988);
{\sl Nucl. Phys.} {\bf B328}, 1 (1989)\\
     V.~W.~Hughes {\it et al.} {\sl Phys.Lett.} {\bf B212}, 511 (1988).

\bibitem{SMC}
     B.~ {\it et al.} {\sl Phys.Lett.} {\bf B302}, 553 (1993);\ 
     {\bf B320}, 400 (1994)\\
     D.~Adams {\it et al.}, {\sl Phys.Lett.} {\bf B329}, 399 (1994).

\bibitem{E142}
     P.~L.~Anthony {\it et al.} {\sl Phys.Rev.Lett.} {\bf 71}, 959 (1993).

\bibitem{E143a}
     K.~Abe {\it et al.} {\sl Phys.Rev.Lett.} {\bf 74}, 346 (1995).

\bibitem{E143b}
     K.~Abe {\it et al.} {\sl Phys.Rev.Lett.} {\bf 75}, 25 (1995).

\bibitem{sv82a}
     E.~V.~Shuryak and A.~I.~Vainshtein, {\sl Nucl. Phys.} 
{\bf B201}, 141 (1982)

\bibitem{jaffe90}
     R.~L.~Jaffe, {\sl Comments Nucl.Part.Phys.} {\bf 19}, 239 (1990)

\bibitem{mit91}
     R.~L.~Jaffe and X. Ji, {\sl Phys. Rev.} {\bf D43}, 724 (1991);

\bibitem{ji94}
     X.~Ji and P.~Unrau, {\sl Phys. Lett.} {\bf B333}, 228 (1994) 

\bibitem{ems94}
     E.~Ehrnsberger, L.~Mankiewicz and A.~Schafer, {\sl Phys.Lett.} 
{\bf B323}, 439 (1994).

\bibitem{jaffe95}
     R.~L.~Jaffe, {\sl Physics Today} September, 24 (1995).

\bibitem{ans94}
     M.~Anselmino, A.~Efremov and E.~Leader, Phys. Rep. {\bf 261}, 
1 (1995)
 
\bibitem{close}
     F.~E.~Close, hep-ph/9509251, September 8 (1995)

\bibitem{hey72}
     A.~J.~G.~Hey and J.~E.~Mandula, {\sl Phys.Rev.} {\bf D5}, 
2610 (1972).

\bibitem{ji93}
     X.~Ji, {\sl Nucl. Phys.} {\bf B402}, 217 (1993)

\bibitem{kod}
     J.~Kodaira, S.~Matsuda, K.~Sasaki and T.~Uematsu, {\sl Nucl.Phys.}
     {\bf B159}, 99 (1979).

\bibitem{ww}
     S.~Wandzura and F.~Wilczek, {\sl Phys.Lett.} {\bf B72}, 195 (1977).

\bibitem{SMC94}
     D.~Adams {\it et al.}, {\sl Phys.Lett.} {\bf B336}, 125 (1994).

\bibitem{E143c}
     K.~Abe {\it et al.} {\sl SLAC-PUB-6982,} Sep. 1, 1995.

\bibitem{song94}
     X.~Song and J.~M.~McCarthy, {\sl Phys.Rev.} {\bf D49}, 3169 (1994).

\bibitem{jaffe95p}
     R.~L.~Jaffe, hep-ph/9509279 (1995)

\bibitem{song92}
     X.~Song and J.~M.~McCarthy, {\sl Phys.Rev.} {\bf C46}, 1077 (1992).

\bibitem{wang95}
     X.~M.~Wang and X.~Song, {\sl Phys.Rev.} {\bf C51}, 2750 (1995).

\bibitem{sst91}
     A.~W.~Schreiber, A.~I.~Signal and A.~W.~Thomas {\sl Phys. Rev.} 
{\bf D44}, 2653 (1991)

\bibitem{strat93}
     M.~Stratmann, {\sl Z. Phys.} {\bf C60}, 763 (1993)

\bibitem{apsong}
     G.~Altarelli and G.~Parisi, {\sl Nucl. Phys.} {\bf B126}, 298 (1977);\\
     X.~Song and J.~Du, {\sl Phys. Rev.} {\bf D40}, 2177 (1989)

\bibitem{sv82b}
     E.~V.~Shuryak and A.~I.~Vainshtein, {\sl Nucl. Phys.} 
{\bf B199}, 451 (1982)

\bibitem{bkl83}
     A.~P.~Bukhvostov, E.~A.~Kureav and L.~N.~Lipatov, {\sl JETP 
Lett.} {\bf 37}, 482 (1983); {\sl Sov. Phys. JETP} {\bf 60}, 22 (1984).

\bibitem{rat86}
     P.~G.~Ratcliffe, {\sl Nucl. Phys.} {\bf B264}, 493 (1986).

\bibitem{bb88}
     I.~I.~Balitsky and V.~M.~Braun, {\sl Nucl. Phys.} {\bf B311}, 
541, (1988/89).

\bibitem{jichou90}
     X.~Ji and C.~Chou, {\sl Phys.Rev.} {\bf D42}, 3637 (1994).

\bibitem{ali91}
     A.~Ali, V.~M.~Braun and G.~Hiller, {\sl Phys.Lett.} {\bf B266}, 
117 (1991).

\bibitem{kyu95}
     A.~Kodaira, Y.~Yasui and T.~Uematsu, {\sl Phys.Lett.} {\bf B344}, 
348 (1995).

\bibitem{tj95}
     T.~J.~Liu, PhD Thesis, Univ. of Virginia, September, 1995 
 
\bibitem{bag95}
     F.~M. Steffens, H.~Holtmann and A.~W.~Thomas, hep-ph/9508398

\bibitem{BC}
      H.~Burkhardt and W.~N.~Cottingham, {\sl Ann. Phys.} {\bf 56}
453 (1970)

\bibitem{qcdsr1}
     E.~Stein {\it et al.}, {\sl Phys. Lett.} {\bf B334}, 369 (1995)

\bibitem{qcdsr2}
     I.~I.~Balitsky, V.~M.~Braun and A.~V.~Kolesnichenko,
{\sl Phys. Lett.} {\bf B242}, 245 (1990); {\bf B318}, 648 (1993)
(Erratum)

\bibitem{lat95}
     M.~Gockeler {\it et al.}, {\sl Phys.Rev.} {\bf D53}, 2317 (1996).

\bibitem{cpr92}
     J.~L.~Cortes, B.~Pire, and J.~P.~Ralston, {\sl Z. Phys.}
{\bf C55},409 (1992) 

\bibitem{E155}
     R.~Arnold {\it et al.} {\sl SLAC-PROPOSAL-E-154,} (1993); 
{\sl SLAC-PROPOSAL-E-155,} (1993). 

\bibitem{SMC94a}
     SMC Collaboration, {\sl Measurements of the Spin-dependent Structure
Functions of the Neutron and the Proton}, Addendum to the NA47 Proposal, 
CERN/SPSLC 94-13 (1994).

\bibitem{hera}
     M.~Dueren and K.~Rith, {\sl Proceedings, Physics at HERA,} 
Vol.{\bf 1}, 427 (1991).


\end{thebibliography}
\end{document}